\pdfoutput=1

\documentclass[journal]{IEEEtran}

\ifCLASSINFOpdf

\else

\fi

\usepackage{xcolor}
\definecolor{bordeau}{rgb}{0.3515625,0,0.234375}
\usepackage[absolute,overlay]{textpos}
\usepackage{graphicx}
\usepackage{amsmath,amssymb,amsfonts, mathtools}  
\usepackage{lipsum}
\usepackage{array}
\usepackage{caption}
\usepackage{multicol}
\usepackage{hyperref}
\usepackage{afterpage}
\usepackage[vlined]{algorithm2e}
\usepackage{algorithmic}
\usepackage{setspace}
\usepackage{tabularx,booktabs}
\usepackage{makecell}
\newcolumntype{C}{>{\centering\arraybackslash}X}
\usepackage{multirow}
\usepackage{caption}
\usepackage{subcaption}
\usepackage{nameref}
\usepackage{pgffor}
\usepackage{colortbl}
\usepackage{hhline}
\usepackage{float}

\usepackage [utf8]{inputenc}
\usepackage [T1] {fontenc}
\definecolor{darkred}{RGB}{142,6,6}
\definecolor{darkblue}{RGB}{37, 65, 109}%{69, 120, 204}
\definecolor{blue}{RGB}{175, 204, 255}
\definecolor{darkgreen}{RGB}{0,78,17}
\definecolor{green}{RGB}{183, 255, 197}%{86,138,97}

\begin{document}

% in *personalized* healthcare?
\title{Integration of Clinical Criteria into the Training of Deep Models: Application to Glucose Prediction for Diabetic People}

\author{Maxime~De~Bois,
        Moun\^{i}m~A.~El~Yacoubi,
        and~Mehdi~Ammi
\thanks{M. De Bois is with CNRS-LIMSI and the Universit\'{e} Paris-Saclay, Orsay, France (e-mail: maxime.debois@limsi.fr).}
\thanks{M. A. El Yacoubi is with Samovar, CNRS, T{\'e}l{\'e}com SudParis, Institut Polytechnique de Paris, \'{E}vry, France}
\thanks{M. Ammi is with Universit\'{e} Paris 8, Saint-Denis, France}}

% \markboth{Journal of \LaTeX\ Class Files,~Vol.~14, No.~8, August~2015}
% {Shell \MakeLowercase{\textit{et al.}}: Bare Demo of IEEEtran.cls for IEEE Journals}

\maketitle

\IEEEpeerreviewmaketitle

\begin{abstract}
Standard objective functions used during the training of neural-network-based predictive models do not consider clinical criteria, leading to models that are not necessarily clinically acceptable. In this study, we look at this problem from the perspective of the forecasting of future glucose values for diabetic people. In this study, we propose the \textit{coherent mean squared glycemic error} (gcMSE) loss function. It penalizes the model during its training not only of the prediction errors, but also on the predicted variation errors which is important in glucose prediction. Moreover, it makes possible to adjust the weighting of the different areas in the error space to better focus on dangerous regions. In order to use the loss function in practice, we propose an algorithm that progressively improves the clinical acceptability of the model, so that we can achieve the best tradeoff possible between accuracy and given clinical criteria. We evaluate the approaches using two diabetes datasets, one having type-1 patients and the other type-2 patients. The results show that using the gcMSE loss function, instead of a standard MSE loss function, improves the clinical acceptability of the models. In particular, the improvements are significant in the hypoglycemia region. We also show that this increased clinical acceptability comes at the cost of a decrease in the average accuracy of the model. Finally, we show that this tradeoff between accuracy and clinical acceptability can be successfully addressed with the proposed algorithm. For given clinical criteria, the algorithm can find the optimal solution that maximizes the accuracy while at the same meeting the criteria.

\end{abstract}

\section{Introduction}

With 4.2 million of imputed deaths in 2019, diabetes is undoubtedly one of the major diseases of our modern world \cite{fid2019atlas}. Compared to healthy persons, diabetic people experience trouble in the regulation of their blood glucose level. Whereas pancreas of type-1 diabetic people do not produce insulin, a hormone responsible for the absorption of glucose in the blood, the body cells of type-2 diabetic patients get increasingly resistant to its action. Failing to regulate the blood sugar level put the patient at risk of getting in states of hypoglycemia and hyperglycemia. In hypoglycemia (blood sugar level below 70 mg/dL) the patient faces short-term consequences such as clumsiness, coma or even death. On the other hand, with hyperglycemia (blood sugar level above 180 mg/dL), the consequences are more long-term with an increased risk of cardiovascular diseases or blindness.

In the recent years, a lot of researchers have been interested in the creation of glucose predictive models \cite{oviedo2017review}. Using past glucose values, carbohydrate (CHO) intakes and insulin infusions information, the models can forecast the future glucose values 30 to 60 minutes ahead of time \cite{oviedo2017review}. For the diabetic patient, being able to know the future values of his/her glycemia could be highly beneficial as hypo/hyperglycemia could be anticipated. Historically, glucose predictive models are based on autoregressive processes \cite{sparacino2007glucose}. However, thanks to the advance in machine learning and deep learning, but also in the increased availability of data, we are currently witnessing a shift in favor of more complex models, and in particular models based on neural networks. The use of standard feedforward neural networks has been explored with, for instance, the works of Pappada \textit{et al.} \cite{pappada2011neural}, Georga \textit{et al.} \cite{georga2013glucose} and Ben Ali \textit{et al.} \cite{ali2018continuous}. Recurrent neural networks, and in particular those based on long short-term memory (LSTM) units, are probably the most popular deep models for glucose prediction. Aliberti \textit{et al.} showed that they are more accurate than standard autoregressive models \cite{aliberti2019multi}. Mirshekarian \textit{et al.} demonstrated their superiority over support vector regression (SVR) models that use expert physiological features \cite{mirshekarian2017using}. They also have been shown to benefit from the addition of various input features such as heart rate or skin conductance \cite{mirshekarian2019lstms,martinsson2019blood}. Other neural-network-based solutions have been tried out recently. Among them, we can highlight the promising use of convolutional neural networks \cite{de2020adversarial, zhu2018deep}.

Models based on neural networks are trained by backpropagating the gradient of the average error to the weights of the network. In glucose prediction, as in almost all regression problems, the average error is computed as the mean squared error (MSE). As a consequence, the models are trained on maximizing the accuracy of the predictions. However, in the benchmark study we recently conducted \cite{de2020glyfe}, we showed that a good accuracy does not ensure that the predictions are clinically acceptable. Indeed, some errors, despite their relatively low magnitude, can be very dangerous for the patient (e.g., errors in the hypoglycemia region). To address this issue, Del Favero \textit{et al.} proposed the gMSE loss function that amplifies the weights of the errors based on the observed glycemic region \cite{del2012glucose}. They showed that using the gMSE instead of the standard MSE decreases the number of dangerous predictions at the cost of reducing the average accuracy of the model. While their methodology is promising, their study has several limitations that we aim at addressing. First, as the approach has been evaluated using autoregressive models on virtual diabetic patients, it is unclear how it translates to more complex models and to real patients. Also, their approach focuses on only one aspect of the clinical acceptability of the predictions, which is the point clinical accuracy. Another aspect of the clinical acceptability of the predictions is the clinical accuracy of predicted variations (i.e., difference between two successive predictions compared to the observed variations), which is taken into account in the widely used \textit{continuous glucose-error grid analysis} (CG-EGA) metric \cite{kovatchev2004evaluating}. Indeed, inaccurate predicted glucose variations can be very dangerous as they can confuse the patient in the understanding of the future evolution of the glycemia.

Our contributions are:
\begin{enumerate}
    \item We propose a new loss function called the \textit{coherent mean squared glycemic error} (gcMSE). Compared to the standard MSE, it includes constaints directly related to the clinical acceptability of the models. In particular, it penalizes the model during its training not only on prediction errors, but also on predicted variations errors \cite{de2019prediction}. Moreover, it makes possible to increase the importance of specific regions in the error space (e.g., the hypoglycemia region).
    \item The gcMSE faces a multi-objective optimization problem. Indeed, when promoting the learning of a model focused more on making clinically acceptable predictions, we reduce the constraints on its global accuracy. However, for the model to be useful for the diabetic patient, it needs to be accurate. To address this challenge, we propose the PICA algorithm that iteratively relax the accuracy constraints so that the focus is progressively more in favor on the satisfaction of the clinical constraints. This enables the creation of a model that maximizes the accuracy while at the same time that respects the given clinical constraints.
    \item We evaluate the proposed solutions on two diabetes datasets, the IDIAB dataset and the OhioT1DM dataset, characterized by their heterogeneity. Whereas the IDIAB dataset, collected by ourselves, is made of 6 type-2 diabetic patients, the OhioT1DM dataset has been released by Marling \textit{et al.} and comprises data from 6 type-1 diabetic patients \cite{marling2018ohiot1dm}.
    \item We open-sourced the code written in Python that has been used in this study in a GitHub repository \cite{debois2018apac}.
\end{enumerate}

The paper is organized as follows. First, after introducing the CG-EGA metric in more details, we present the whole framework for its integration into the training of deep models. Then, we describe the machine learning pipeline, with the preprocessing of the data, the models we used, and the evaluation process. Finaly, we present and discuss the experimental results.

\section{Integrating Clinical Criteria into the Training of Deep Models}

In this section we propose a method to integrate the clinical criteria of the CG-EGA within the training of deep models. First, we introduce, in details, how the CG-EGA metric works. Then, we present the gcMSE loss function that integrates the clinical constraints. Finally, we propose a methodology to use this new cost function in practice.

\subsection{Presentation of the CG-EGA}

\begin{figure*}[!ht]
    \centering
    \includegraphics{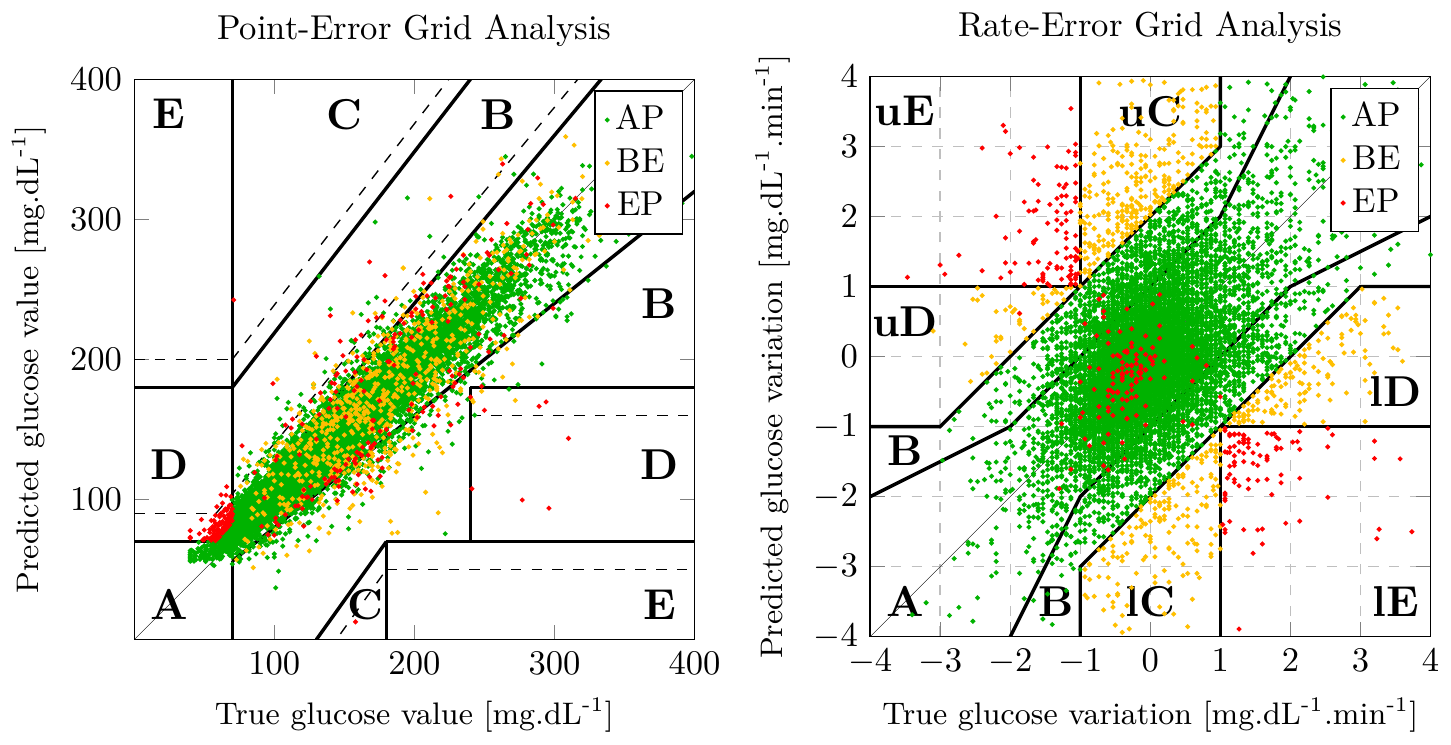}
    \caption{Example of the CG-EGA classification with the P-EGA (left) and R-EGA (right).}
    \label{fig:cgega}
\end{figure*}

Originally proposed by Kovatchev \textit{et al.} for the evaluation of the clinical acceptability of blood glucose sensors \cite{kovatchev2004evaluating}, the \textit{continuous glucose-error grid analysis} (CG-EGA) is a widely used metric to assess the clinical acceptability of glucose predictive models \cite{oviedo2017review}. It is made of the combination of two different evaluation grids: the \textit{point-error grid analysis} (P-EGA) and the \textit{rate-error grid analysis} (R-EGA). While the P-EGA measure the clinical accuracy of the predictions, the R-EGA measures the clinical accuracy of the predicted variations. The predicted variations are computed as the rate of change between two consecutive predictions. Both grids attribute to a given prediction a score from A (best) to E (worst) representing the dangerousness of the prediction. Figure \ref{fig:cgega} gives a graphical representation of both grids. The scores in both grids are then combined into a final label assessing the clinical acceptability of the metric. A prediction can either be an \textit{accurate prediction} (AP), a \textit{benign error} (BE), or an \textit{erroneous prediction} (EP).

Table \ref{table:cgega} details the reasoning behind the CG-EGA scores. First, the CG-EGA has a different behavior depending on the glycemic region (hypoglycemia, euglycemia, or hyperglycemia). Essentially, the glycemic region impacts the way bad R-EGA scores (C to E) are accounted. For instance, in the hypoglycemia region, a lE score in the R-EGA, representing a fast predicted decrease in glycemia while a fast increase is observed, can lead to a benign error (BE) if the last prediction is accurate (A in the P-EGA). In the hypoglycemia region, the CG-EGA implies that it is not dangerous for the patient to predict a decrease in glycemia as it will not lead to life-threatening actions from the patient. On the other hand, the absence of detection of negative variations in the uD and uE zones is extremely dangerous: hypoglycemia is becoming much worse, which could result in consequences such as coma or even death. Overall, for a prediction to be labelled as an accurate prediction AP, it needs good scores (A or B) in both grids.

\definecolor{darkred}{RGB}{142,6,6}
\definecolor{red}{RGB}{255, 183, 183}%{183,74,74}
\definecolor{darkblue}{RGB}{37, 65, 109}%{69, 120, 204}
\definecolor{blue}{RGB}{175, 204, 255}
\definecolor{darkgreen}{RGB}{0,78,17}
\definecolor{green}{RGB}{183, 255, 197}%{86,138,97}

\begin{table*}[!ht]
\caption{Classification of glucose predictions performed by the CG-EGA. Depending on the scores obtained on the P-EGA and R-EGA, a prediction is classified as an accurate prediction (AP), a benign error (BE) or erroneous prediction (EP).}
\footnotesize
\setlength\extrarowheight{4pt}
\label{table:cgega}
% \begin{tabularx}{\textwidth}{@{}c|c|c|c|c|c|c|c|c|c|c|c|c|@{}}
\begin{center}
\begin{tabular}{@{}|c|c|c|c|c|c|c|c|c|c|c|c|c|@{}}

\hline

\multicolumn{2}{|c|}{\multirow{3}{*}{}} &    \multicolumn{11}{c|}{\textbf{P-EGA}} \\

\cline{3-13}

\multicolumn{2}{|c|}{} & \multicolumn{3}{c|}{\textit{Hypoglycemia}} & \multicolumn{3}{c|}{\textit{Euglycemia}} & \multicolumn{5}{c|}{\textit{Hyperglycemia}} \\ 

% \cellcolor{green} \color{darkgreen} AP
% \cellcolor{blue} \color{darkblue} BE
% \cellcolor{red} \color{darkred} EP

\cline{3-13}
\multicolumn{2}{|c|}{} & A & D & E & A & B & C & A & B & C & D & E \\
\cline{1-13}
\parbox[t]{2mm}{\multirow{8}{*}{\rotatebox[origin=c]{90}{\textbf{R-EGA}}}} & A & \cellcolor{green} \color{darkgreen} AP & \cellcolor{red} \color{darkred} EP & \cellcolor{red} \color{darkred} EP & \cellcolor{green} \color{darkgreen} AP & \cellcolor{green} \color{darkgreen} AP & 
\cellcolor{red} \color{darkred} EP& \cellcolor{green} \color{darkgreen} AP & \cellcolor{green} \color{darkgreen} AP & \cellcolor{red} \color{darkred} EP & \cellcolor{red} \color{darkred} EP& \cellcolor{red} \color{darkred} EP\\
\cline{2-13}%\hhline{~~|-|-|-|-|-|-|-|-|-|-|-|}
% \cline{2-13}
& B & \cellcolor{green} \color{darkgreen} AP & \cellcolor{red} \color{darkred} EP & \cellcolor{red} \color{darkred} EP & \cellcolor{green} \color{darkgreen} AP & \cellcolor{green} \color{darkgreen} AP & 
\cellcolor{red} \color{darkred} EP& \cellcolor{green} \color{darkgreen} AP & \cellcolor{green} \color{darkgreen} AP & \cellcolor{red} \color{darkred} EP & \cellcolor{red} \color{darkred} EP& \cellcolor{red} \color{darkred} EP\\
\cline{2-13}
& uC & \cellcolor{blue} \color{darkblue} BE & \cellcolor{red} \color{darkred} EP & \cellcolor{red} \color{darkred} EP & \cellcolor{blue} \color{darkblue} BE & \cellcolor{blue} \color{darkblue} BE & 
\cellcolor{red} \color{darkred} EP& \cellcolor{blue} \color{darkblue} BE & \cellcolor{blue} \color{darkblue} BE & \cellcolor{red} \color{darkred} EP & \cellcolor{red} \color{darkred} EP& \cellcolor{red} \color{darkred} EP\\
\cline{2-13}
& lC & \cellcolor{blue} \color{darkblue} BE & \cellcolor{red} \color{darkred} EP & \cellcolor{red} \color{darkred} EP & \cellcolor{blue} \color{darkblue} BE & \cellcolor{blue} \color{darkblue} BE & 
\cellcolor{red} \color{darkred} EP& \cellcolor{blue} \color{darkblue} BE & \cellcolor{blue} \color{darkblue} BE & \cellcolor{red} \color{darkred} EP & \cellcolor{red} \color{darkred} EP& \cellcolor{red} \color{darkred} EP\\
\cline{2-13}
& uD & \cellcolor{red} \color{darkred} EP & \cellcolor{red} \color{darkred} EP & \cellcolor{red} \color{darkred} EP & \cellcolor{blue} \color{darkblue} BE & \cellcolor{blue} \color{darkblue} BE & 
\cellcolor{red} \color{darkred} EP& \cellcolor{blue} \color{darkblue} BE & \cellcolor{blue} \color{darkblue} BE & \cellcolor{red} \color{darkred} EP & \cellcolor{red} \color{darkred} EP& \cellcolor{red} \color{darkred} EP\\
\cline{2-13}
& lD & \cellcolor{blue} \color{darkblue} BE & \cellcolor{red} \color{darkred} EP & \cellcolor{red} \color{darkred} EP & \cellcolor{blue} \color{darkblue} BE & \cellcolor{blue} \color{darkblue} BE & 
\cellcolor{red} \color{darkred} EP& \cellcolor{red} \color{darkred} EP & \cellcolor{red} \color{darkred} EP & \cellcolor{red} \color{darkred} EP & \cellcolor{red} \color{darkred} EP& \cellcolor{red} \color{darkred} EP\\
\cline{2-13}
& uE & \cellcolor{red} \color{darkred} EP & \cellcolor{red} \color{darkred} EP & \cellcolor{red} \color{darkred} EP & \cellcolor{red} \color{darkred} EP & \cellcolor{red} \color{darkred} EP & 
\cellcolor{red} \color{darkred} EP& \cellcolor{red} \color{darkred} EP & \cellcolor{red} \color{darkred} EP & \cellcolor{red} \color{darkred} EP & \cellcolor{red} \color{darkred} EP& \cellcolor{red} \color{darkred} EP\\
\cline{2-13}
& lE & \cellcolor{blue} \color{darkblue} BE & \cellcolor{red} \color{darkred} EP & \cellcolor{red} \color{darkred} EP & \cellcolor{red} \color{darkred} EP & \cellcolor{red} \color{darkred} EP & 
\cellcolor{red} \color{darkred} EP& \cellcolor{red} \color{darkred} EP & \cellcolor{red} \color{darkred} EP & \cellcolor{red} \color{darkred} EP & \cellcolor{red} \color{darkred} EP& \cellcolor{red} \color{darkred} EP\\

\hline

\end{tabular}

~

AP: Accurate Prediction; BE: Benign Error; EP: Erroneous Prediction

\end{center}

% \begin{flushright}
% \end{flushright}

\end{table*}

In summary, compared to standard accuracy metrics such as the \textit{root mean squared error} (RMSE), the CG-EGA also evaluates the accuracy of the predicted variations. And, most importantly, these evaluations depend on the observed glycemic region. These aspects should be taken into account if we want to add clinical constraints based on the CG-EGA into the training of the models.

\subsection{Coherent Mean Squared Error}

In deep learning, the models are trained by backpropagating the gradient of the loss function to the networks' weights. Thus, by modifying the objective function, it is possible to modify the predictive behavior of the model. We can find numerous cost functions in the literature, the most used being the cross-entropy for classification problems and the \textit{mean squared error} (MSE) for regression problems. Since glucose prediction is a regression task, deep models in the field use the MSE in the model's training. Equation \ref{eqn:mse} describes MSE as the squared difference between the observed $\boldsymbol{g}$ and predicted $\boldsymbol{\hat{g}}$, averaged over $N$ samples. In this study, we propose modifications to the MSE cost function to improve the clinical acceptability of the predictions.

\begin{equation}
    MSE(\boldsymbol{g},\boldsymbol{\hat{{g}}})=\frac{1}{N} \sum_{n=1}^N ({g}_{n}-\hat{{g}}_{n})^2
\label{eqn:mse}    
\end{equation}

First, as we have seen by analyzing the CG-EGA behavior, it is essential to penalize predicted variation errors in addition to prediction errors. To do this, we can use the coherent mean squared error (cMSE) loss function, previously proposed in a work of ours \cite{de2019prediction}. The cMSE is the MSE of the predictions weighted by the MSE of the predicted variations. The Equation \ref{eqn:cmse} describes the cMSE loss function with $\Delta{\boldsymbol{g}}$ and $\Delta{\boldsymbol{\hat{g}}}$ representing, respectively, the observed and predicted glucose variations. We call the weighting coefficient $c$ the coherence factor. It represents the relative importance we give to the accuracy of the predicted variations versus the accuracy of the predictions.

\begin{equation} \label{eqn:cmse}
\begin{split}
cMSE(\boldsymbol{g},\boldsymbol{\hat{g}}) & = MSE(\boldsymbol{g},\boldsymbol{\hat{g}}) + c \cdot MSE(\boldsymbol{\Delta g},\boldsymbol{\Delta \hat{g}}) \\
 & = \frac{1}{N} \sum_{n=1}^N ({g}_n-\hat{{g}}_n)^2 + c \cdot (\Delta{g}_n-\Delta\hat{{g}}_n)^2
\end{split}
\end{equation}

\begin{figure*}[!ht]
\centering
\includegraphics [width=0.65\linewidth]{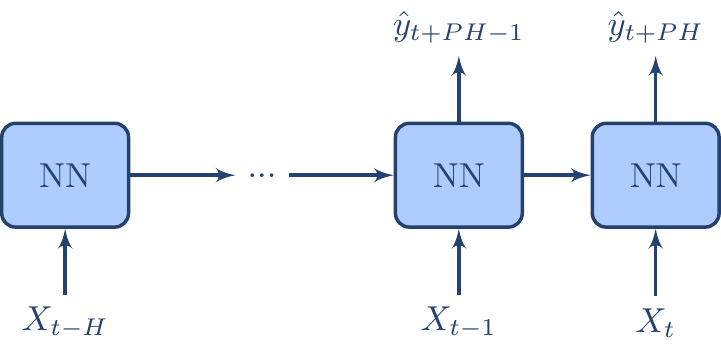}  
\caption{General architecture of a two-output recurrent neural network that has been unrolled $H$ times, where $H$ is the length of the history of input data to the model. $X_t$ are the input data to the model at time $t$ (e.g., glucose, insulin, and carbohydrates at time $t$), and $\hat{y}_{t+PH}$ is the model prediction (e.g., blood glucose prediction) at $t+PH$, where $PH$ is the prediction horizon.}
\label{fig:two-output-rnn}
\end{figure*}

To be able to use the cMSE, we can use a recurrent neural network (e.g., LSTM) with two outputs (see Figure \ref{fig:two-output-rnn}). The two outputs represent the prediction at the given prediction horizon $PH$ and the prediction at $PH - \Delta T$, $\Delta T$ being the time interval between two predictions. For instance, with a prediction interval of 5 minutes and a prediction horizon of 30 minutes, the networks outputs the predictions at the horizons 30 and 25 minutes. These two outputs enables the computation of the predicted variations, as depicted by Equation \ref{eqn:deltag}. The architecture of recurrent neural networks is particularly suited to this task since each sub-module of the unfolded network (see Figure \ref{fig:two-output-rnn}) shares the same weights.

\begin{equation}
    \Delta{\hat{g}}_{t+PH} = \frac{\hat{g}_{t+PH} - \hat{g}_{t+PH-\Delta T}}{\Delta T}
\label{eqn:deltag}    
\end{equation}

\subsection{Coherent Mean Squared Glycemic Error}

The analysis of the CG-EGA showed us that glucose prediction errors and predicted variation errors do not have the same clinical importance in the error space (see Table \ref{table:cgega}). Although generally of greater magnitude, these clinical errors are rare and represent only a small portion of the gradient in the updating of the network's weights during its training. Therefore, minimizing the MSE (or, equivalently, the cMSE) does not directly reduce the number of clinical errors. Indeed, most of the weight updates are focused towards the improvement of the accuracy of predictions that already have good a clinical acceptability. In the field of multi-class classification, it is very common to weight samples from under-represented classes by artificially increasing their presence within the training set. In their work on object recognition within images, Lin \textit{et al.} proposed to dynamically weight the learning samples according to their difficulty (a sample being considered easy when the probability of the corresponding class is very high, showing a high degree of confidence in the model) \cite{lin2017focal}. By reducing the weights of samples judged easy, the training of the model focuses on the samples for which it has the most difficulty. Finally, Del Favero \textit{et al.} proposed, in the context of glucose prediction, to modify the MSE to better account for the dangerous regions of the P-EGA \cite{del2012glucose}. In particular, they proposed that samples with observed hypoglycemia or hyperglycemia should be given a higher weighting. Although this work was evaluated on autoregressive models and virtual patients, their results showed that this new cost function reduces the number of predictions in zone D and E of the P-EGA grid. 

Taking inspiration from their work, we propose to dynamically penalize prediction errors as well as predicted variation errors. This new cost function, named \textit{coherent mean squared glycemic error} (gcMSE), penalizes predictions differently depending on the P-EGA and R-EGA regions (see Equation \ref{eqn:gcMSE}). In Equation \ref{eqn:gcMSE-P}, $P_X$ and $p_x$, $X \in \{A,B,uC,lC,uD,lD,uE,lE\}$ and $x \in \{a,b,uC,lC,uD,ld,uE,le\}$, represent the P-EGA grid regions and their respective weights. Contrary to the original P-EGA, we have segmented the C, D and E regions in two, as it is already the case for the R-EGA. This allows more flexibility in assigning weights. Equivalently, in Equation \ref{eqn:gcMSE-R}, $R_X$ and $r_x$, $X \in \{A,B,uC,lC,uD,lD,uE,lE\}$ and $x \in \{a,b,uC,lC,uD,ld,uE,le\}$ represent the regions of the R-EGA grid and their respective weights.

\begin{subequations}\label{eqn:gcMSE}
\begin{subequations}
% \begin{alignat}{2}
% gcMSE(\boldsymbol{g},\boldsymbol{\hat{g}}) =& P(\boldsymbol{g},\boldsymbol{\hat{g}}) \cdot MSE(\boldsymbol{g},\boldsymbol{\hat{g}}) \\
% & + c \cdot R(\boldsymbol{\Delta g},\boldsymbol{\Delta \hat{g}}) \cdot MSE(\boldsymbol{\Delta g},\boldsymbol{\Delta \hat{g}})  \tag{\ref{eqn:gcMSE}}
% \end{alignat}
\begin{multline}
gcMSE(\boldsymbol{g},\boldsymbol{\hat{g}}) = P(\boldsymbol{g},\boldsymbol{\hat{g}}) \cdot MSE(\boldsymbol{g},\boldsymbol{\hat{g}}) \\
+ c \cdot R(\boldsymbol{\Delta g},\boldsymbol{\Delta \hat{g}}) \cdot MSE(\boldsymbol{\Delta g},\boldsymbol{\Delta \hat{g}})  \tag{\ref{eqn:gcMSE}}
\end{multline}
\end{subequations}
with,
\begin{equation}\label{eqn:gcMSE-P}
P(\boldsymbol{g},\boldsymbol{\hat{g}}) = 
\begin{cases}
      p_a, & \text{if}\ \{\boldsymbol{g},\boldsymbol{\hat{g}}\} \in P_A \\
      p_b, & \text{if}\ \{\boldsymbol{g},\boldsymbol{\hat{g}}\} \in P_B \\
      p_{uc}, & \text{if}\ \{\boldsymbol{g},\boldsymbol{\hat{g}}\} \in P_{uC} \\
      p_{lc}, & \text{if}\ \{\boldsymbol{g},\boldsymbol{\hat{g}}\} \in P_{lC} \\
      p_{ud}, & \text{if}\ \{\boldsymbol{g},\boldsymbol{\hat{g}}\} \in P_{uD} \\
      p_{ld}, & \text{if}\ \{\boldsymbol{g},\boldsymbol{\hat{g}}\} \in P_{lD} \\
      p_{ue}, & \text{if}\ \{\boldsymbol{g},\boldsymbol{\hat{g}}\} \in P_{uE} \\
      p_{le}, & \text{if}\ \{\boldsymbol{g},\boldsymbol{\hat{g}}\} \in P_{lE} \\
    \end{cases}
\end{equation}
and,
\begin{equation}\label{eqn:gcMSE-R}
R(\boldsymbol{\Delta g},\boldsymbol{\Delta \hat{g}}) =
\begin{cases}
      r_a, & \text{if}\ \{\boldsymbol{\Delta g},\boldsymbol{\Delta \hat{g}}\} \in R_A \\
      r_b, & \text{if}\ \{\boldsymbol{\Delta g},\boldsymbol{\Delta \hat{g}}\} \in R_B \\
      r_{uc}, & \text{if}\ \{\boldsymbol{\Delta g},\boldsymbol{\Delta \hat{g}}\} \in R_{uC} \\
      r_{lc}, & \text{if}\ \{\boldsymbol{\Delta g},\boldsymbol{\Delta \hat{g}}\} \in R_{lC} \\
      r_{ud}, & \text{if}\ \{\boldsymbol{\Delta g},\boldsymbol{\Delta \hat{g}}\} \in R_{uD} \\
      r_{ld}, & \text{if}\ \{\boldsymbol{\Delta g},\boldsymbol{\Delta \hat{g}}\} \in R_{lD} \\
      r_{ue}, & \text{if}\ \{\boldsymbol{\Delta g},\boldsymbol{\Delta \hat{g}}\} \in R_{uE} \\
      r_{le}, & \text{if}\ \{\boldsymbol{\Delta g},\boldsymbol{\Delta \hat{g}}\} \in R_{lE} \\
    \end{cases}
\end{equation}
\end{subequations}

Using the gcMSE instead of the standard MSE introduces 14 new hyperparameters to be optimized: the $c$ coherence factor, and the weights associated with the P-EGA and R-EGA regions. This task being particularly laborious, we propose simplifications reducing the number of hyperparameters :

\begin{itemize}
    \item First, it is not interesting to improve the accuracy of the predicted variations in zones A and B. Indeed, all predictions belonging to these zones are clinically sufficiently accurate. Thus, we can set $r_a=r_b=0$.
    \item From the perspective of the possible maximization of the AP rate, BE and EP predictions can be seen as equally important. This allows us to set most of the C, D and E zones to the same value. Moreover, the coherence factor $c$ alone allows us to weight the importance we give between the accuracy of the predictions and the accuracy of predicted variations. Thus, we can decide to set all these weights to 1.
    \item Only the hypoglycemic P-EGA regions D and E ($P_{uD}$ and $P_{uE}$) require a special treatment in order to increase the importance of samples in the hypoglycemic region. We denote the weight associated with these areas by $p_{hypo}$.
\end{itemize}

 The Equation \ref{eqn:gcMSE-simple} summarizes these design simplifications, allowing the gcMSE cost function to have only 3 hyperparameters: $p_{ab}$, $p_{hypo}$, and $c$. The choice of these hyperparameters depends on both the learning objectives and the experimental conditions. The coherence factor $c$ must be chosen depending on the importance of the cost function $MSE(\boldsymbol{\Delta g},\boldsymbol{\Delta \hat{g}})$ compared to the $MSE(\boldsymbol{g},\boldsymbol{\hat{g}})$. The choice of the coefficient $p_{hypo}$ must be made according to the size of the datasets. When few hypoglycemic samples are available, it is possible to give a value of $p_{hypo} > 1$. As for $p_{ab}$, it represents the precision constraint we give during training. The lower its value, the more the training of the model focuses on improving its clinical acceptability at the expense of its accuracy.

\begin{subequations}\label{eqn:gcMSE-simple}
\begin{equation}
P(\boldsymbol{g},\boldsymbol{\hat{g}}) = 
\begin{cases}
      p_{ab}, & \text{si}\ \{\boldsymbol{g},\boldsymbol{\hat{g}}\} \in \{P_A,P_B\} \\
      p_{hypo}, & \text{si}\ \{\boldsymbol{g},\boldsymbol{\hat{g}}\} \in \{P_{uD}, P_{uE}\} \\
      1, & \text{else}\\
    \end{cases}
\end{equation}
and,
\begin{equation}
R(\boldsymbol{\Delta g},\boldsymbol{\Delta \hat{g}}) =
\begin{cases}
      0, & \text{si}\ \{\boldsymbol{\Delta g},\boldsymbol{\Delta \hat{g}}\} \in \{R_A,R_B\} \\
      1, & \text{else}\\
    \end{cases}
\end{equation}
\end{subequations}

\subsection{Progressive Improvement of the Clinical Acceptability}

In order to be able to use the gcMSE cost function, we need to formulate the learning objective, and in particular the relative importance of improving the clinical acceptability. Indeed, as shown in the work of Del Favero \textit{et al.}, an improvement in the clinical acceptability is often matched by a deterioration in the statistical accuracy \cite{del2012glucose}.

Research in the field of multi-objective optimization (MOO) highlights the need of using selection criteria, which can take the form of a weighting between the different objectives, or thresholds for the different objectives \cite{marler2004survey}. Even though there is no standard clinical criterion for glucose prediction models today, we propose to project ourselves by assuming their existence. These clinical criteria could take the form of minimum thresholds in AP and/or maximum thresholds in EP following the CG-EGA (e.g., minimum 95\% of predictions obtaining the AP score in to CG-EGA). Our learning objective in this case would be to maximize the accuracy of the predictions while meeting the clinical criteria.

To achieve this goal, we need to test a large number of different model architectures (hyperparameters), each test involving the training of a neural network. This training is very expensive in the context of deep learning. Therefore, an efficient training methodology must be used in order to reach the optimal solution. The methodologies generally used to answer multi-objective optimization problems are based on genetic methods (such as NSGA-II \cite{deb2000fast}). Although faster than a simple grid search, these algorithms involve randomness in the changes made to the different tests slowing down the convergence.

In order to circumvent this problem, we propose the \textit{progressive improvement of clinical acceptability} (PICA) algorithm. Starting from a solution that maximizes the model's accuracy without taking into account its clinical acceptability, the constraints on the precision are gradually relaxed in favor of its clinical acceptability. This has the consequence of gradually degrading the statistical accuracy of the model, a degradation that is accompanied by a progressive improvement in clinical acceptability.

\RestyleAlgo{ruled}
\LinesNumbered
\begin{algorithm}
 \KwData{clinical criteria C, model M, update coefficient $\alpha$, smoothing coefficient $\beta$}
 \KwResult{Model maximizing the accuracy while respecting the clinical criteria C or $-1$}

 $i \leftarrow 0$
 
 $M_0 \leftarrow $ train(MSE)
 
 $\boldsymbol{g}_0, \hat{\boldsymbol{g}}_0 \leftarrow$ predict($M_0$)
 
 $\hat{\boldsymbol{g}}_0^* \leftarrow$ smooth($\hat{\boldsymbol{g}}_0,\beta)$
 
  \While{$C(M_i) = False$ et $\text{MASE}(\boldsymbol{g}_i, \hat{\boldsymbol{g}}_i^*) < 1$}{
 $i \leftarrow i + 1$ 
 
$\text{gcMSE}_i \leftarrow \text{gcMSE}$ avec $p_{ab} \leftarrow \alpha ^ {i - 1}$
 
 $M_i \leftarrow $ finetune($M_0$, cMSE\textsubscript{i})
 
 $\boldsymbol{g}_i, \hat{\boldsymbol{g}}_i \leftarrow$ predict($M_i$)
 
 $\hat{\boldsymbol{g}}_i^* \leftarrow$ smooth($\hat{\boldsymbol{g}}_i,\beta)$
 
 }
 
 \If{$\text{MASE}(\boldsymbol{g}_i, \hat{\boldsymbol{g}}_i^*) < 1$}{ \KwRet{$M_i$}}
 \Else{\KwRet{$-1$}}
 
 \caption{Progressive Improvement of the Clinical Acceptability (PICA)}
 \label{algo:APAC}
\end{algorithm}

The Algorithm \ref{algo:APAC} gives a description of the steps of the PICA algorithm. The updating law of the weights $p_{ab}$, representing the constraints in the statistical accuracy, is to be chosen according to the experimental conditions. In this study, we use the law defined by the Equation \ref{eqn:updaterule} (with $\alpha \in [0, 1]$ being the speed of the relaxation of the accuracy constraints). As for the MASE metric (\textit{mean absolute scaled error}, proposed by Hyndman \textit{et al.} \cite{hyndman2006another}, see Equation \ref{eqn:mase}), it is used as a stopping criterion when clinical criteria are not achievable. The algorithm stops when the MASE exceeds 1, meaning that a naïve prediction model (whose prediction is equal to the last known observation) is more accurate than the present model. Finally, we use an exponential smoothing of the predictions. This smoothing allows to attenuate the important fluctuations of the predictions in the first steps of the algorithm. By being small, it allows a significant gain in clinical acceptability, in return for a minimal loss of accuracy. For more details on the exponential smoothing of the predictions, we invite the reader to refer to the post-processing steps in Section \ref{sec:cgega_postprocessing}.

\begin{equation}
    \label{eqn:updaterule}
    p_{ab} = \alpha ^ {i - 1}
\end{equation}

\begin{equation}
    MASE(\boldsymbol{g},\boldsymbol{\hat{{g}}}, PH)=\frac{\frac{1}{N} \cdot \sum_{n=1}^{N} \left| g_n - \hat{g}_n \right|}{\frac{1}{N-PH} \cdot \sum_{n=PH}^{N} \left| g_n - g_{n-PH} \right|} 
\label{eqn:mase}    
\end{equation}

The PICA algorithm avoids unnecessary iterations, each iteration bringing the model closer and closer to its goal. Moreover, instead of being trained from its initial state, the model is refined from the first model, trained with the standard MSE. This refinement requires much less iteration than a full training, and thus allows the algorithm to run faster. Another approach would have been to refine the model from the previous iteration instead. However, in practice, we were confronted with a local optimum problem, preventing the model from finding a better solution after updating the cost function.

\section{Methods}

In this section, we present the whole methodology that has been followed for the evaluation of the proposed losses and the PICA algorithm. First, we present the experimental datasets and their preprocessing. Then,  we provide details about the post-processing of the predictions and the models' evaluation. Finally, we describe the different models with their implementation.

We have made the whole implementation of the data pipeline available in a GitHub repository \cite{debois2018apac}.

\subsection{Experimental Data}

In this study, we used two datasets made of several diabetic patients: the IDIAB dataset and the OhioT1DM dataset. While the IDIAB has been collected by ourselves between 2018 and 2019 after the approval by the French ethical commitee (ID RCB 2018-A00312-53), the OhioT1DM has recently been released by Marling \textit{et al.} \cite{marling2018ohiot1dm}.

~
\subsubsection{IDIAB Dataset (I)} The IDIAB dataset is made of 6 type-2 diabetic patients (5F/1M, age 56.5 $\pm$ 9.14 years old, BMI 33.52 $\pm$ 4.17 $kg/m^2$). The patients had been monitored for 31.17 $\pm$ 1.86 days in free-living conditions. We collected glucose values (in mg/dL) by using FreeStyle Libre continuous glucose monitoring devices (Abbott Diabetes Care). As for carbohydrate (CHO) intakes (g) and insulin infusion values (unit), they have been manually recorded with the mySugr coaching application for diabetes.

~
\subsubsection{OhioT1DM Dataset (O)} The OhioT1DM is made of data coming from 6 type-1 diabetic patients (2M/4F, age between 40 and 60 years old, BMI not disclosed) that had been monitored for 8 weeks in free living conditions. For more information concerning the experimental system, we redirect the reader to \cite{marling2018ohiot1dm}. We restrict ourselves to the glucose values, the insulin infusions, and the CHO intakes to remain consistent with IDIAB data.

\subsection{Preprocessing}

\begin{figure*}[ht]
	\includegraphics[width=\textwidth]{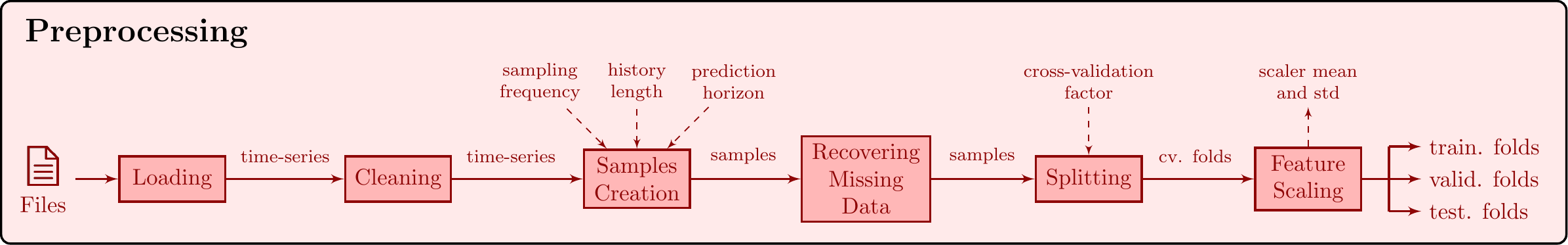}
	\centering
	\caption{Preprocessing of the data.}
	\centering
	\label{fig:preprocessing}
\end{figure*}

The preprocessing stage aims at preparing the data for their use in the training and the evaluation of the models. It is made of several steps depicted by Figure \ref{fig:preprocessing} and described in the following paragraphs.

~
\subsubsection{Cleaning}

The glucose time-series from the IDIAB dataset possess several erroneous values. These values are characterized by peaks lasting only one sample. We decided to remove these samples from the data as keeping them would be hurtful for the training as well for the evaluation of the models. Instead of removing them by hand, we used an automated methodology proposed in our previous work \cite{de2019prediction}. A sample is flagged as erroneous if the surrounding rates of change are incoherent with the typical distribution of rates of change, and if they are of opposite signs.

~
\subsubsection{Samples Creation}

The two datasets have been resampled to a sample every 5 minutes which is the sampling frequency of the OhioT1DM glucose signal. While we took the mean of the glucose signals, the CHO and insulin values have been accumulated.

The input samples have been obtained by using a sliding window of length $H$ of 3 hours (36 samples) on the three signals. The prediction objective is, for each sample, the glucose value 30 minutes (6 samples) in the future (prediciton horizon, PH, of 30 minutes).

~
\subsubsection{Recovering Missing Data}

Both datasets contain numerous missing values coming either from sensor or human errors. Moreover, contrary to the OhioT1DM dataset, the upsampling of the IDIAB glucose signal (from 15 minutes to 5 minutes) has introduced a lot of missing values as well. We can artificially recover some of them by following the following strategy for every sample:
\begin{enumerate}
    \item linearly interpolate the glucose history when the missing value is surrounded by two known glucose values;
    \item extrapolate linearly in the opposite case, usually when the missing glucose value is the most recent data;
    \item discard samples when the ground truth $y_{t+PH}$ is not known to prevent training and testing on artificial data.
\end{enumerate}

~
\subsubsection{Splitting}

The datasets are split into training, validation and testing sets. While the testing set is used for the final evaluation of the models, the validation is used as a prior evaluation for the optimization of the models' hyperparameters.

The testing set is made of the last 10 days for the OhioT1DM dataset and of the last 5 days for the IDIAB dataset, the latter being around two times smaller. The remaining days have been split into training and validation sets following an 80\%/20\% distribution with 5 permutations.

~
\subsubsection{Feature Scaling}

Finally, the samples have been standardized (zero mean and unit variance) w.r.t. their training set.

\subsection{Post-processing and Evaluation}

\label{sec:cgega_postprocessing}

\begin{figure*}[!ht]
	\includegraphics[width=\textwidth]{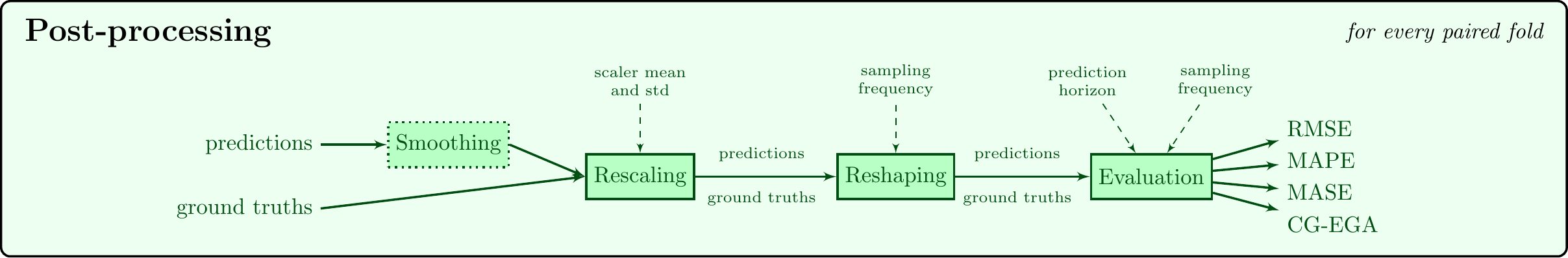}
	\centering
	\caption{Post-processing and evaluation of the predictions.}
	\centering
	\label{fig:retain_postprocessing}
\end{figure*}

The evaluation of the predictive models is done following the steps described by Figure \ref{fig:retain_postprocessing}. In this study, we focus models that are personalized to the patient and that predict future glucose values with a 30-minute prediction horizon. Before evaluating the predictions, we follow two mandatory post-processing steps. First, we rescale the predictions to their original scale (see the \textit{features scaling} preprocessing step). Then, we reconstruct the prediction time-series by reordering the predictions. In addition, the predictions made by the models can be smoothed, as it is done in the PICA algorithm.

~
\subsubsection{Exponential Smoothing}

The PICA algorithm involves the smoothing of the predictions at each iteration. The objective of this smoothing is to reduce excessive fluctuations in the predicted glucose signal. These oscillations are not representative of actual glucose variations and are therefore dangerous for the patient.

We chose the exponential smoothing technique rather than the moving average technique because it gives more weight to recent predictions. Exponential smoothing can be defined as recursive, with each value of the smoothed signal being equal to a weighting between the value of the original signal and the previous value of the smoothed signal (see Equation \ref{eqn:exponential_smoothing}, where $\hat{g}_t^*$ represents the smoothed value of the glucose prediction $\hat{g}_t$ and $\beta$ the smoothing coefficient) \cite{brown2004smoothing}.

\begin{equation}
    \hat{g}^*_t = \begin{cases}
      \hat{g}_0, & \text{if}\ t=0 \\
      \beta \cdot \hat{g}_t + (1 - \beta) \cdot \hat{g}^*_{t-1}, & \text{else}\\
    \end{cases}
    \label{eqn:exponential_smoothing}
\end{equation}

The higher $\beta$ is, the stronger is the weight given to the original signal, and the less smooth the signal is. The choice of the $\beta$ smoothing coefficient in $[0, 1] $ must be made carefully.  Indeed, too aggressive smoothing will result in a temporal shift of the signal. In the context of glucose prediction, this will greatly reduce the accuracy of the model, and therefore its usefulness for the patient.

To our knowledge, although common in signal processing (e.g., power consumption prediction \cite{taylor2007short}), no post-processing smoothing has been done in the literature of glucose prediction. We can nevertheless note the occasional use of low-pass filters (which act similarly to the exponential smoothing technique) on the input signal \cite{sparacino2007glucose,perez2010artificial}.

~
\subsection{Metrics}

To evaluate the models we use four different metrics: the RMSE, the MAPE, the MASE and the CG-EGA. For each metric, the performances are averaged over the 5 test subsets of each patient linked to a 5-fold cross-validation on the training/validation permutations. They are then also averaged on all the patients from the same dataset. Both the RMSE, MAPE and MASE metrics give a complementary measure of the accuracy of the prediction. While the RMSE is closely related to the prediction scale, the MAPE is scale independent and is expressed in percentage. As for the MASE, it measures the average usefulness of the predictions compared to naïve predictions (predictions equal to the last known observations). The MASE is computed following Equation \ref{eqn:mase}, presented in the previous section. On the other hand, the CG-EGA measures the clinical acceptability of the prediction by analyzing the clinical accuracy as well as the coherence between successive predictions. In the end, the CG-EGA classifies a prediction either as an accurate prediction (AP), a benign error (BE), or an erroneous prediction (EP). A high AP rate and a low EP rate are necessary for a model to be clinically acceptable. The rates can be either averaged over all the test samples, or for the samples within a specific glycemic region (i.e., hypoglycemia, euglycemia and hyperglycemia).

\subsection{Glucose Predictive Models}

The objective of the study is to improve the clinical acceptability of deep models. To this end, we first proposed a new cost function cMSE which penalizes the model during its training not only on prediction errors but also on predicted variation errors. We then proposed the gcMSE, which is the cMSE customized to glucose prediction. In particular, it introduces weighting coefficients based on the CG-EGA to enhance the clinical acceptability of the model. Finally, we proposed the PICA algorithm that allows to progressively improve the clinical acceptability of the models through the use of the gcMSE function. The models that we present here aim at evaluating these different proposals.

We use as reference models the Support Vector Regression model (SVR) and Long Short-Term Memory recurrent neural network (LSTM) from the GLYFE benchmark study \cite{debois2018GLYFE}. As the preprocessing steps are identical in the two studies, the results are fully comparable. The SVR and LSTM models represent, respectively, the best model and the best deep model in this benchmark.

First, to analyze the potential improvement of the clinical acceptability through the cMSE and gcMSE cost functions, we can evaluate the pcLSTM and gpcLSTM models respectively. These two models are based on a two-output LSTM architecture, which, apart from the presence of the two outputs, is identical to the LSTM model of the GLYFE benchmark study. They are respectively trained to minimize the cMSE and gcMSE loss functions with a coherence factor $c$ set to 8 for the IDIAB dataset and 2 for the OhioT1DM dataset. This difference between the two sets is explained by a MSE of the predicted variations being approximately 4 times greater for the OhioT1DM dataset. As for the coefficients $p_{ab}$ and $p_{hypo}$ of the gcMSE, we set them to 1 and 10 respectively. These coefficients are identical to those from the first iteration of the PICA algorithm. In addition, we propose to evaluate an additional variant of the gcMSE whose coefficient $p_{ab}$ is set to 0. This model, denoted gpcLSTM$_\text{CA}$, is a model that aims at maximizing the clinical acceptability, without taking into account the precision of the model beyond clinical acceptability needs.

The PICA algorithm uses the exponential smoothing technique to stabilize successive predictions. In order to fully evaluate the impact of the cost functions and the PICA algorithm, we use the exponential smoothing technique on all the models presented in this study. The smoothed variant of each model is represented by a superscript asterisk (e.g., LSTM$^*$, pcLSTM$^*$, gpcLSTM$^*_\text{CA}$). All these models use a smoothing coefficient of 0.85, as it degrades only slightly the accuracy of the predicted signal.

The PICA algorithm makes a compromise between the gpcLSTM$^*$ and gpcLSTM$^*_\text{CA}$ models. The emphasis on clinical acceptability of this compromise is progressive over the iterations of the algorithm. However, the precision constraint, through the coefficient $p_{ab}$ is never equal to 0 (model gpcLSTM$^*_\text{CA}$), because such a model has a precision far too low to be useful for the diabetic patient. This is why the PICA algorithm stops when the MASE exceeds the value of 1 on the validation set. We represent by the model gpcLSTM$^*_\text{PICA}$ the results obtained when the PICA algorithm stops. These results present the upper bounds of clinical acceptability while maintaining a useful accuracy. In the PICA algorithm, we use the coefficient $p_{ab}$ update law presented by Equation \ref{eqn:updaterule}. It involves the coefficient $\alpha$, the rate at which the constraint in accuracy is relaxed, which has been set to 0.9 in this study. A higher coefficient gives better control over the final trade-off, in return for a slower execution time (more iterations before convergence). The PICA algorithm uses exponential smoothing on the model's predictions to increase the stability of the predicted signal. The smoothing coefficient $\beta$, as for all the smoothed variants of the other models, it was fixed at 0.85.

\section{Experimental Results}

\subsection{Presentation of the Experimental Results}

\begin{table*}[!ht]
\footnotesize
\centering
\caption{Mean (with standard deviation) of statistical accuracy (RMSE, MAPE, and MASE) and general clinical acceptability (CG-EGA) for a prediction horizon of 30 minutes and for the IDIAB and OhioT1DM datasets.}
\label{table:cgega_general_results}
\begin{center}

\begin{tabular}{c||c|c|c|ccc}
\toprule

\multirow{2}{*}{\textbf{Model}} &\multirow{2}{*}{\textbf{RMSE}} &\multirow{2}{*}{\textbf{MAPE}} &\multirow{2}{*}{\textbf{MASE}} &   \multicolumn{3}{c}{\textbf{CG-EGA (general)}} \\

&&&& AP & BE & EP \\

\toprule

\multicolumn{7}{c}{\textbf{\textit{IDIAB Dataset}}}\\
\midrule
\textbf{SVR} & 20.32 \scriptsize{(6.02)} & \underline{\textbf{8.66 \scriptsize{(0.44)}}} & 0.85 \scriptsize{(0.15)} & 92.69 \scriptsize{(2.81)} & 5.34 \scriptsize{(2.06)} & 1.97 \scriptsize{(1.23)}\\

\textbf{LSTM} & \underline{\textbf{19.85 \scriptsize{(6.00)}}} & 9.04 \scriptsize{(1.11)} & \underline{\textbf{0.85 \scriptsize{(0.10)}}}  & 92.20 \scriptsize{(2.99)} & 5.05 \scriptsize{(1.71)} & 2.76 \scriptsize{(1.82)}\\

\textbf{SVR\textsuperscript{*}} & 20.67 \scriptsize{(6.20)} & 8.86 \scriptsize{(0.44)} & 0.88 \scriptsize{(0.15)} & 93.62 \scriptsize{(2.57)} & 4.47 \scriptsize{(1.69)} & 1.92 \scriptsize{(1.35)}\\

\textbf{LSTM\textsuperscript{*}} & 20.27 \scriptsize{(6.30)} & 9.25 \scriptsize{(1.21)} & 0.87 \scriptsize{(0.09)}  & 93.16 \scriptsize{(3.13)} & 4.16 \scriptsize{(1.75)} & 2.68 \scriptsize{(2.00)}\\

\midrule

\textbf{pcLSTM} & 21.89 \scriptsize{(5.68)} & 10.28 \scriptsize{(1.34)} & 0.96 \scriptsize{(0.11)} & 94.04 \scriptsize{(3.26)} & 3.20 \scriptsize{(1.66)} & 2.76 \scriptsize{(2.07)}\\

\textbf{pcLSTM$^\text{*}$} & 22.63 \scriptsize{(6.04)} & 10.64 \scriptsize{(1.40)} & 1.00 \scriptsize{(0.11)} & 94.24 \scriptsize{(3.35)} & \underline{\textbf{2.94 \scriptsize{(1.73)}}} & 2.82 \scriptsize{(2.07)}\\

\textbf{gpcLSTM} & 21.21 \scriptsize{(5.64)} & 9.35 \scriptsize{(0.92)} & 0.91 \scriptsize{(0.13)} & 94.03 \scriptsize{(2.66)} & 3.91 \scriptsize{(1.48)} & 2.06 \scriptsize{(1.54)}\\
\textbf{gpcLSTM$^\text{*}$} & 21.86 \scriptsize{(5.94)} & 9.66 \scriptsize{(0.95)} & 0.94 \scriptsize{(0.13)} & 94.53 \scriptsize{(2.84)} & 3.38 \scriptsize{(1.55)} & 2.08 \scriptsize{(1.57)}\\

\textbf{gpcLSTM$_{\text{CA}}$} & 40.68 \scriptsize{(11.20)} & 18.14 \scriptsize{(5.55)} & 1.91 \scriptsize{(0.55)} & 95.34 \scriptsize{(2.76)} & 3.29 \scriptsize{(2.56)} & \underline{\textbf{1.37 \scriptsize{(0.91)}}}\\
\textbf{gpcLSTM$_{\text{CA}}^{\text{*}}$} & 41.15 \scriptsize{(11.18)} & 18.36 \scriptsize{(5.47)} & 1.93 \scriptsize{(0.54)} & \underline{\textbf{95.35 \scriptsize{(2.87)}}} & 3.20 \scriptsize{(2.61)} & 1.45 \scriptsize{(0.92)}\\

\textbf{gpcLSTM$^\text{*}_{\text{PICA}}$} & 24.03 \scriptsize{(7.15)} & 10.43 \scriptsize{(1.18)} & 1.03 \scriptsize{(0.09)} & 95.00 \scriptsize{(2.74)} & 3.38 \scriptsize{(1.99)} & 1.61 \scriptsize{(1.22)}\\

\midrule
\multicolumn{7}{c}{\textbf{\textit{OhioT1DM Dataset}}}\\
\midrule
\textbf{SVR} & \underline{\textbf{20.15 \scriptsize{(2.33)}}} & \underline{\textbf{9.12 \scriptsize{(2.11)}}} & \underline{\textbf{0.85 \scriptsize{(0.02)}}} & 83.35 \scriptsize{(3.91)} & 12.38 \scriptsize{(2.83)} & 4.28 \scriptsize{(1.83)}\\

\textbf{LSTM} & 20.46 \scriptsize{(2.08)} & 9.24 \scriptsize{(2.10)} & 0.86 \scriptsize{(0.02)} & 80.03 \scriptsize{(4.17)} & 14.83 \scriptsize{(2.88)} & 5.14 \scriptsize{(2.11)}\\

\textbf{SVR\textsuperscript{*}} & 20.17 \scriptsize{(2.30)} & 9.18 \scriptsize{(2.12)} & \underline{\textbf{0.85 \scriptsize{(0.02)}}} & 85.00 \scriptsize{(4.05)} & 10.97 \scriptsize{(2.72)} & 4.03 \scriptsize{(1.90)}\\

\textbf{LSTM\textsuperscript{*}} & 20.43 \scriptsize{(2.03)} & 9.26 \scriptsize{(2.10)} & 0.86 \scriptsize{(0.02)} & 82.14 \scriptsize{(3.94)} & 13.06 \scriptsize{(2.51)} & 4.81 \scriptsize{(2.04)}\\

\midrule

\textbf{pcLSTM} & 21.53 \scriptsize{(2.23)} & 10.07 \scriptsize{(2.32)} & 0.93 \scriptsize{(0.03)} & 87.45 \scriptsize{(3.76)} & 8.46 \scriptsize{(2.05)} & 4.09 \scriptsize{(2.14)}\\

\textbf{pcLSTM$^\text{*}$} & 21.71 \scriptsize{(2.22)} & 10.19 \scriptsize{(2.35)} & 0.94 \scriptsize{(0.03)} & 87.89 \scriptsize{(3.61)} & 8.15 \scriptsize{(1.94)} & 3.96 \scriptsize{(2.12)}\\

\textbf{gpcLSTM} & 21.66 \scriptsize{(2.69)} & 9.65 \scriptsize{(2.14)} & 0.92 \scriptsize{(0.03)} & 86.97 \scriptsize{(3.63)} & 9.50 \scriptsize{(2.52)} & 3.53 \scriptsize{(1.48)}\\
\textbf{gpcLSTM$^\text{*}$} & 21.82 \scriptsize{(2.69)} & 9.76 \scriptsize{(2.16)} & 0.93 \scriptsize{(0.03)} & 87.59 \scriptsize{(3.45)} & 9.01 \scriptsize{(2.31)} & 3.41 \scriptsize{(1.49)}\\

\textbf{gpcLSTM$_{\text{CA}}$} & 47.70 \scriptsize{(6.31)} & 22.43 \scriptsize{(2.76)} & 2.37 \scriptsize{(0.53)} & 90.46 \scriptsize{(2.85)} & 7.16 \scriptsize{(1.66)} & \underline{\textbf{2.37 \scriptsize{(1.28)}}}\\

\textbf{gpcLSTM$_{\text{CA}}^{\text{*}}$} & 47.82 \scriptsize{(6.27)} & 22.47 \scriptsize{(2.76)} & 2.37 \scriptsize{(0.53)} & \underline{\textbf{90.51 \scriptsize{(2.88)}}} & \underline{\textbf{7.12 \scriptsize{(1.64)}}} & {{2.37 \scriptsize{(1.30)}}}\\

\textbf{gpcLSTM$^\text{*}_{\text{PICA}}$} & 23.50 \scriptsize{(2.49)} & 10.46 \scriptsize{(2.09)} & 1.01 \scriptsize{(0.03)} & 88.72 \scriptsize{(3.59)} & 8.20 \scriptsize{(2.23)} & 3.08 \scriptsize{(1.64)}\\

\bottomrule
\end{tabular}

\end{center}
\end{table*}{}
\begin{table*}[!ht]
\caption{Mean (with standard deviation) of per-region clinical acceptability (CG-EGA) for a prediction horizon of 30 minutes and for the IDIAB and OhioT1DM datasets.}
\footnotesize
\label{table:cgega_cgega_results}
\begin{tabularx}{\linewidth}{@{}c||*{3}{c}|*{3}{c}|*{3}{c}@{}}
\toprule

\multirow{3}{*}{\textbf{Model}} &    \multicolumn{9}{c}{\textbf{CG-EGA (per region)}} \\

& \multicolumn{3}{c|}{\textit{Hypoglycemia}} & \multicolumn{3}{c|}{\textit{Euglycemia}} & \multicolumn{3}{c}{\textit{Hyperglycemia}} \\ 

& AP & BE & EP & AP & BE & EP & AP & BE & EP \\

\midrule
\multicolumn{10}{c}{\textbf{\textit{IDIAB Dataset}}}\\
\midrule
\textbf{SVR} & 69.39 \scriptsize{(33.51)} & 0.35 \scriptsize{(0.70)} & 30.27 \scriptsize{(33.54)} & 95.17 \scriptsize{(2.01)} & 4.33 \scriptsize{(1.83)} & 0.50 \scriptsize{(0.47)} & 89.51 \scriptsize{(6.09)} & 7.43 \scriptsize{(3.86)} & 3.06 \scriptsize{(2.53)}\\

\textbf{LSTM} & 40.94 \scriptsize{(30.73)} & \underline{\textbf{0.00 \scriptsize{(0.00)}}} & 59.06 \scriptsize{(30.73)} & 95.78 \scriptsize{(1.48)} & 3.83 \scriptsize{(1.55)} & 0.39 \scriptsize{(0.38)} & 89.55 \scriptsize{(5.60)} & 7.35 \scriptsize{(3.21)} & 3.10 \scriptsize{(2.45)}\\

\textbf{SVR\textsuperscript{*}} & 66.37 \scriptsize{(31.47)} & 0.17 \scriptsize{(0.35)} & 33.45 \scriptsize{(31.51)} & 96.13 \scriptsize{(1.81)} & 3.49 \scriptsize{(1.66)} & 0.39 \scriptsize{(0.36)} & 90.61 \scriptsize{(5.67)} & 6.60 \scriptsize{(3.23)} & 2.79 \scriptsize{(2.79)}\\

\textbf{LSTM\textsuperscript{*}} & 37.99 \scriptsize{(31.22)} & \underline{\textbf{0.00 \scriptsize{(0.00)}}} & 62.01 \scriptsize{(31.22)} & 96.71 \scriptsize{(1.35)} & 2.95 \scriptsize{(1.46)} & 0.33 \scriptsize{(0.38)} & 91.02 \scriptsize{(6.04)} & 6.18 \scriptsize{(3.67)} & 2.80 \scriptsize{(2.58)}\\

\midrule

\textbf{pcLSTM} & 34.59 \scriptsize{(29.27)} & \underline{\textbf{0.00 \scriptsize{(0.00)}}} & 65.41 \scriptsize{(29.27)} & 97.58 \scriptsize{(0.90)} & 2.13 \scriptsize{(0.82)} & 0.29 \scriptsize{(0.20)} & 92.60 \scriptsize{(5.81)} & 4.94 \scriptsize{(3.18)} & 2.46 \scriptsize{(2.80)}\\
\textbf{pcLSTM$^\text{*}$}  & 32.20 \scriptsize{(27.83)} & \underline{\textbf{0.00 \scriptsize{(0.00)}}} & 67.80 \scriptsize{(27.83)} & \underline{\textbf{97.96 \scriptsize{(0.98)}}} & \underline{\textbf{1.81 \scriptsize{(0.91)}}} & \underline{\textbf{0.23 \scriptsize{(0.11)}}} & 92.81 \scriptsize{(6.25)} & \underline{\textbf{4.68 \scriptsize{(3.48)}}} & 2.51 \scriptsize{(2.85)}\\
\textbf{gpcLSTM} & 64.79 \scriptsize{(24.95)} & \underline{\textbf{0.00 \scriptsize{(0.00)}}} & 35.21 \scriptsize{(24.95)} & 96.60 \scriptsize{(1.11)} & 3.03 \scriptsize{(0.99)} & 0.37 \scriptsize{(0.26)} & 92.06 \scriptsize{(5.12)} & 5.42 \scriptsize{(2.83)} & 2.51 \scriptsize{(2.46)}\\
\textbf{gpcLSTM$^\text{*}$}& 61.87 \scriptsize{(25.17)} & \underline{\textbf{0.00 \scriptsize{(0.00)}}} & 38.13 \scriptsize{(25.17)} & 97.23 \scriptsize{(1.17)} & 2.46 \scriptsize{(1.02)} & 0.31 \scriptsize{(0.22)} & 92.65 \scriptsize{(5.60)} & 4.85 \scriptsize{(3.09)} & 2.50 \scriptsize{(2.68)}\\

\textbf{gpcLSTM$_{\text{CA}}$}& \underline{\textbf{87.95 \scriptsize{(9.58)}}} & 1.71 \scriptsize{(3.43)} & \underline{\textbf{10.34 \scriptsize{(8.15)}}} & 97.37 \scriptsize{(1.36)} & 2.12 \scriptsize{(1.03)} & 0.51 \scriptsize{(0.40)} & 92.17 \scriptsize{(4.46)} & 5.11 \scriptsize{(4.52)} & 2.72 \scriptsize{(2.39)}\\
\textbf{gpcLSTM$_{\text{CA}}^{\text{*}}$} & 87.77 \scriptsize{(9.53)} & 1.71 \scriptsize{(3.43)} & 10.51 \scriptsize{(8.13)} & 97.50 \scriptsize{(1.32)} & 1.97 \scriptsize{(0.97)} & 0.52 \scriptsize{(0.44)} & 92.10 \scriptsize{(4.69)} & 5.03 \scriptsize{(4.70)} & 2.87 \scriptsize{(2.33)}\\

\textbf{gpcLSTM$^\text{*}_{\text{PICA}}$} & 68.49 \scriptsize{(27.85)} & 0.57 \scriptsize{(1.14)} & 30.94 \scriptsize{(28.22)} & 97.35 \scriptsize{(1.18)} & 2.32 \scriptsize{(1.08)} & 0.33 \scriptsize{(0.15)} & \underline{\textbf{93.16 \scriptsize{(4.84)}}} & 5.08 \scriptsize{(3.53)} & \underline{\textbf{1.76 \scriptsize{(1.49)}}}\\

\midrule
\multicolumn{10}{c}{\textbf{\textit{OhioT1DM Dataset}}}\\
\midrule

\textbf{SVR} & 49.71 \scriptsize{(18.75)} & 5.62 \scriptsize{(4.02)} & 44.67 \scriptsize{(18.70)} & 86.35 \scriptsize{(4.24)} & 10.71 \scriptsize{(3.26)} & 2.94 \scriptsize{(1.23)} & 80.85 \scriptsize{(3.24)} & 14.77 \scriptsize{(3.01)} & 4.37 \scriptsize{(1.84)}\\

\textbf{LSTM} & 38.37 \scriptsize{(23.17)} & 3.97 \scriptsize{(3.72)} & 57.67 \scriptsize{(24.23)} & 83.78 \scriptsize{(5.33)} & 12.70 \scriptsize{(4.06)} & 3.52 \scriptsize{(1.47)} & 76.86 \scriptsize{(3.70)} & 17.87 \scriptsize{(2.73)} & 5.27 \scriptsize{(2.21)}\\

\textbf{SVR\textsuperscript{*}} & 46.95 \scriptsize{(21.11)} & 5.97 \scriptsize{(4.05)} & 47.09 \scriptsize{(21.65)} & 87.83 \scriptsize{(4.22)} & 9.46 \scriptsize{(3.21)} & 2.71 \scriptsize{(1.22)} & 82.81 \scriptsize{(3.43)} & 13.12 \scriptsize{(2.98)} & 4.07 \scriptsize{(2.00)}\\

\textbf{LSTM\textsuperscript{*}} & 37.34 \scriptsize{(23.50)} & 4.11 \scriptsize{(4.15)} & 58.56 \scriptsize{(24.17)} & 85.71 \scriptsize{(4.83)} & 11.10 \scriptsize{(3.58)} & 3.19 \scriptsize{(1.37)} & 79.27 \scriptsize{(3.55)} & 15.85 \scriptsize{(2.40)} & 4.88 \scriptsize{(2.24)}\\

\midrule

\textbf{pcLSTM}& 25.28 \scriptsize{(19.11)} & 3.64 \scriptsize{(3.73)} & 71.08 \scriptsize{(19.35)} & 90.79 \scriptsize{(3.43)} & 6.93 \scriptsize{(2.53)} & 2.28 \scriptsize{(1.01)} & 85.78 \scriptsize{(3.64)} & 10.83 \scriptsize{(2.55)} & 3.40 \scriptsize{(2.03)}\\

\textbf{pcLSTM$^\text{*}$} & 23.82 \scriptsize{(18.23)} & 3.72 \scriptsize{(3.48)} & 72.45 \scriptsize{(18.55)} & 91.20 \scriptsize{(3.17)} & 6.67 \scriptsize{(2.35)} & 2.13 \scriptsize{(0.96)} & 86.33 \scriptsize{(3.54)} & 10.44 \scriptsize{(2.50)} & 3.23 \scriptsize{(1.96)}\\

\textbf{gpcLSTM}  & 53.66 \scriptsize{(22.59)} & 4.34 \scriptsize{(3.83)} & 42.00 \scriptsize{(22.86)} & 89.39 \scriptsize{(3.91)} & 7.99 \scriptsize{(2.90)} & 2.63 \scriptsize{(1.12)} & 84.61 \scriptsize{(3.84)} & 11.79 \scriptsize{(3.20)} & 3.61 \scriptsize{(2.01)}\\

\textbf{gpcLSTM$^\text{*}$} & 52.37 \scriptsize{(22.06)} & 4.32 \scriptsize{(3.15)} & 43.30 \scriptsize{(22.42)} & 90.02 \scriptsize{(3.69)} & 7.47 \scriptsize{(2.77)} & 2.52 \scriptsize{(1.04)} & 85.27 \scriptsize{(3.69)} & 11.31 \scriptsize{(2.95)} & 3.42 \scriptsize{(2.02)}\\
\textbf{gpcLSTM$_{\text{CA}}$} & \underline{\textbf{91.17 \scriptsize{(8.50)}}} & 1.26 \scriptsize{(2.08)} & \underline{\textbf{7.57 \scriptsize{(8.01)}}} & 91.61 \scriptsize{(2.03)} & 6.62 \scriptsize{(1.39)} & 1.77 \scriptsize{(0.74)} & \underline{\textbf{87.97 \scriptsize{(5.00)}}} & \underline{\textbf{8.67 \scriptsize{(2.64)}}} & \underline{\textbf{3.36 \scriptsize{(2.63)}}}\\

\textbf{gpcLSTM$_{\text{CA}}^{\text{*}}$}  & 91.02 \scriptsize{(8.49)} & \underline{\textbf{1.21 \scriptsize{(1.97)}}} & 7.77 \scriptsize{(8.00)} & \underline{\textbf{91.71 \scriptsize{(2.02)}}} & \underline{\textbf{6.55 \scriptsize{(1.34)}}} & \underline{\textbf{1.75 \scriptsize{(0.77)}}} & 87.95 \scriptsize{(5.05)} & 8.69 \scriptsize{(2.69)} & \underline{\textbf{3.36 \scriptsize{(2.62)}}}\\

\textbf{gpcLSTM$^\text{*}_{\text{PICA}}$} & 61.30 \scriptsize{(20.12)} & 2.92 \scriptsize{(2.38)} & 35.79 \scriptsize{(20.23)} & 90.84 \scriptsize{(3.57)} & 7.04 \scriptsize{(2.57)} & 2.11 \scriptsize{(1.07)} & 86.48 \scriptsize{(3.95)} & 10.07 \scriptsize{(2.66)} & 3.45 \scriptsize{(2.31)}\\

\bottomrule
\end{tabularx}
\end{table*}

In this section we present the experimental results of this study. These results are represented in the form of two tables: Table \ref{table:cgega_general_results} and \ref{table:cgega_cgega_results}. While Table \ref{table:cgega_general_results} describes the general results of the different models in terms of RMSE, MAPE, MASE and general CG-EGA, Table \ref{table:cgega_cgega_results} gives a more detailed description, by region, of the CG-EGA.

Within our two reference models, SVR and LSTM, the SVR model is the model with the best clinical acceptability (general or regional CG-EGA) for comparable accuracy. In particular, the SVR model has one of the best clinical acceptability in the hypoglycemia region (69.39\% and 49.71\% AP for the IDIAB and OhioT1DM datasets respectively). The exponential smoothing improves the clinical acceptability of the SVR model (SVR$^\text{*}$ model) by -12.79\%\footnote{Here we represent the decrease, in \%, of what is metrically improvable. For the AP, which has a maximum of 100\%, the ratio of change is calculated as $(100 - AP_1) / (100 - AP_2)$.} of AP rate for an increase of +0.90\% in RMSE (decrease in accuracy). The LSTM$^\text{*}$ model is subject to similar changes with -11.44\% AP and +0.98\% RMSE. Table \ref{table:cgega_cgega_results} shows that these improvements in clinical acceptability occur in the euglycemia or hyperglycemia regions, and not in the hypoglycemia region (small decrease in AP).

The pcLSTM model and its smoothed variant pcLSTM$^\text{*}$, using the cMSE cost function as well as the two-output architecture of the LSTM network, are showed to improve the clinical acceptability while deteriorating the accuracy. In particular, the pcLSTM$^\text{*}$ model compared to the LSTM$^\text{*}$ model has -24.18\% AP, and +8.95\% RMSE. The improvement in clinical acceptability is greater for the OhioT1DM dataset (-32.19\% AP) than for the IDIAB dataset (-16.16\% AP). For a comparable decrease in accuracy, the OhioT1DM set benefits more from the cMSE cost function than the IDIAB set. Moreover, the pcLSTM$^\text{*}$ model has among the best clinical acceptability scores in the euglycemia and hyperglycemia regions. However, in comparison with the LSTM or LSTM$^\text{*}$ models, the clinical acceptability in the hypoglycemia region is deteriorated, especially for the OhioT1DM dataset.

The gpcLSTM and gpcLSTM$^\text{*}$ models, using the gcMSE cost function, cMSE customized to blood glucose prediction, show a degradation of the RMSE and an improvement of the AP rate similar to the pcLSTM and pcLSTM$^\text{*}$ models. However, the gpcLSTM and gpcLSTM$^\text{*}$ models have a lower EP rate (-19.53\% and -20.07\% respectively), suggesting an improved clinical acceptability. Table \ref{table:cgega_cgega_results} shows that this improvement is mainly in the hypoglycemia region with much lower EP rates.

The models gpcLSTM$_\text{CA}$ and gpcLSTM$^\text{*}_\text{CA}$ use a gcMSE function with the coefficient $p_{ab}$ of 0. Thus, these models focus only on improving the clinical acceptability. Not seeking to improve the accuracy of predictions beyond the required clinical accuracy (P-EGA Zone B), these models have a very poor RMSE, MAPE and MASE. Nevertheless, they have the best clinical acceptability, with the highest AP and the lowest EP rates. The improvement is particularly important in the hypoglycemic region, as can be seen in Table \ref{table:cgega_cgega_results}.

The gpcLSTM$^\text{*}_\text{PICA}$ model represents the latest iteration of the PICA algorithm with a MASE on the validation set of less than 1. This model is intended to maximize the clinical acceptability, while having a reasonable accuracy (MASE less than 1). Compared to the gpcLSTM$^\text{*}_\text{CA}$ model, it has a slightly lower clinical acceptability (but better than all other models, thanks in particular to its low EP rate).

\subsection{Discussion}

The results show us that exponential smoothing reduces the benign error (BE) rate in favor of a better AP rate, by reducing the amplitude of the variations between successive predictions. This improvement is valid for most of the models and has for counterpart a rather small decrease in the general accuracy of the model. Thus, exponential smoothing, used softly (coefficient $\beta$ of 0.85) is an efficient method to improve the stability of the prediction signal, making it safer for the diabetic patient. However, it remains useless in the hypoglycemia range where the majority of clinical prediction errors are due to poor accuracy.

The effects of using the cMSE cost function on glucose predictions are similar: successive glucose predictions are more consistent with each other, resulting in a large reduction in the BE rate. The effects are greater for the OhioT1DM dataset, which sees its EP rate decrease at the same time. We can explain this by a higher noise in the predicted glucose signal of the OhioT1DM set, noise that comes from the initial glucose signal. With its lower sampling frequency, the IDIAB glucose signal manages to be less noisy in comparison. The cMSE allows successive predictions to be made with a rate of change that better reflects the actual rate of change and thus improves its clinical acceptability. However, like exponential smoothing, improvements in clinical acceptability are not generalized to all glycemic regions. In particular, the hypoglycemic region appears to suffer from the use of cMSE with an increase in its EP rate, especially for the OhioT1DM dataset.

\begin{figure*}[!ht]
        \centering
        \includegraphics[width=\textwidth]{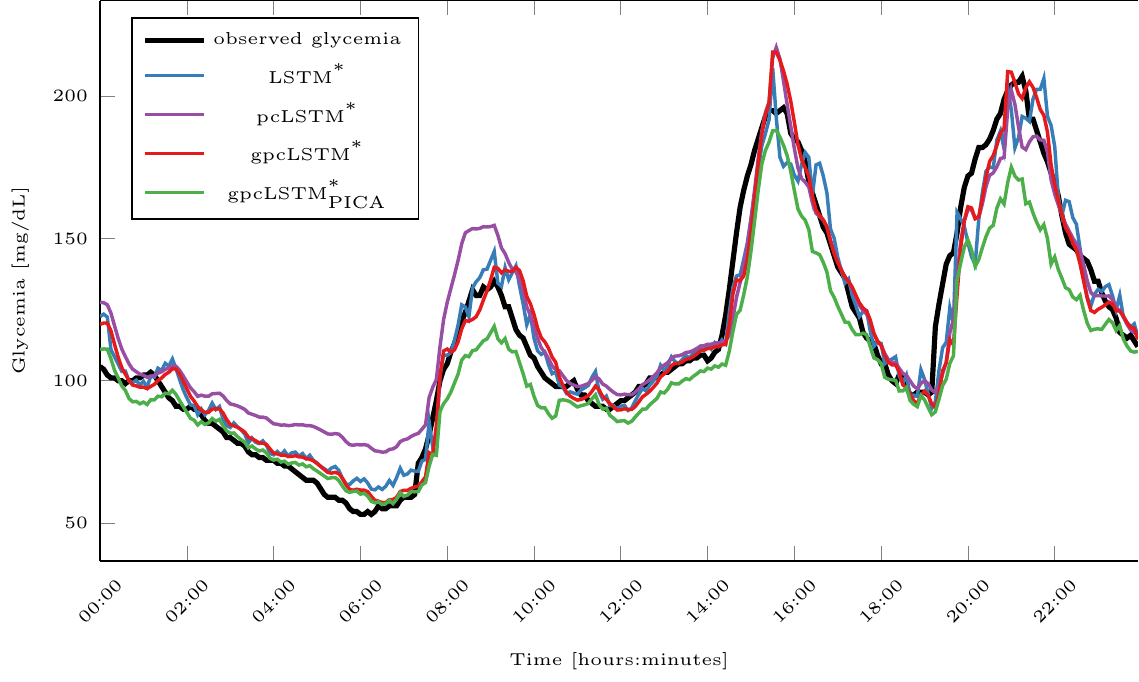}
        \caption{Predictions of the LSTM$^\text{*}$, pcLSTM$^\text{*}$, gpcLSTM$^\text{*}$ and gpcLSTM$^\text{*}_\text{PICA}$ models for the patient 575 from the OhioT1DM dataset for a given day.}  
        \label{fig:cgega_whole_day_prediction}
\end{figure*}

The gcMSE action is more focused on the decrease of the EP rate, as shown by the models gpcLSTM, gpcLSTM$_\text{CA}$, gpcLSTM$_\text{PICA}^\text{*}$. In contrast with the exponential smoothing technique and the cMSE cost function, the gcMSE improves all glycemic regions, and in particular the hypoglycemic region. Moreover, these improvements allow the LSTM neural network to surpass, in clinical acceptability, the SVR model which is the best model of the GLYFE benchmark study. Figure \ref{fig:cgega_whole_day_prediction} allows us to appreciate the differences in the predictions of the different models. First, we can see the large variations and noise in the predicted glucose signal of the LSTM model. These oscillations are reduced for the other models, becoming closer to the observed glucose signal. However, when using the cMSE cost function (pcLSTM$^\text{*}$ signal in purple), we witness a large loss of accuracy in the hypoglycemia region (between 4:00 and 8:00 am). While the signal gpcLSTM$^\text{*}_\text{PICA}$ shows to be very close to the signal observed in the hypoglycemia region, this is done at the cost of an overall loss in accuracy. Finally, gpcLSTM$^\text{*}$, is a compromise between the two.

\begin{figure*}
        \centering
        \begin{subfigure}[b]{0.45\textwidth}
            \centering
            \includegraphics{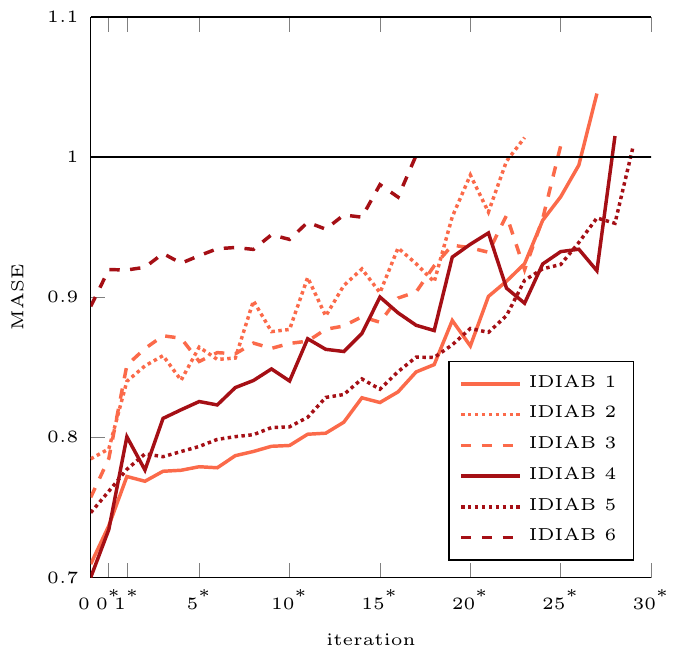}
            \caption[Network1]%
            {{\small MASE evolution of the IDIAB patients}}    
            \label{fig:cg_ega_algo_mase_idiab}
        \end{subfigure}
        \hfill
        \begin{subfigure}[b]{0.45\textwidth}
            \centering
            \includegraphics{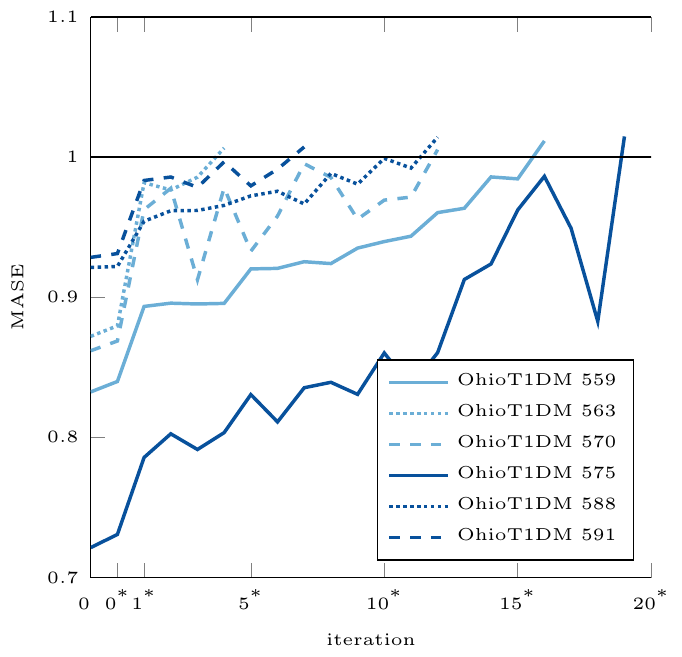}
            \caption[Network1]%
            {{\small MASE evolution of the OhioT1DM patients}}    
            \label{fig:cg_ega_algo_mase_ohio}
        \end{subfigure}
        \vskip\baselineskip
        \begin{subfigure}[b]{0.45\textwidth}  
            \centering 
            \includegraphics{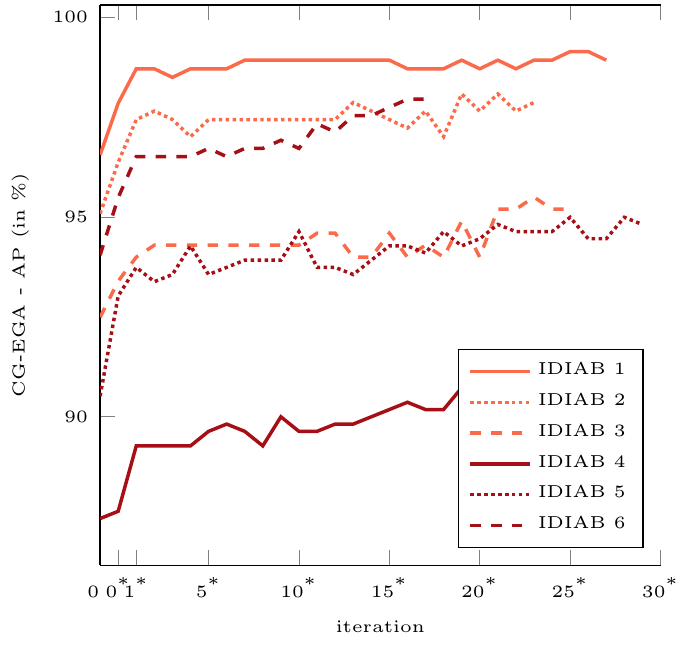}
            \caption[]%
            {{\small AP evolution of the IDIAB patients}}    
            \label{fig:cg_ega_algo_ap_idiab}
        \end{subfigure}
        \hfill
        \begin{subfigure}[b]{0.45\textwidth}  
            \centering 
            \includegraphics{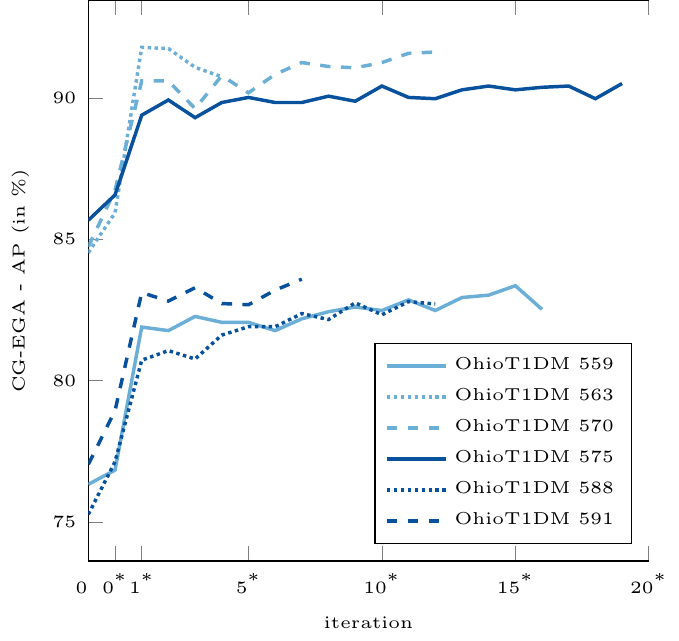}
            \caption[]%
            {{\small AP evolution of the OhioT1DM patients}}    
            \label{fig:cg_ega_algo_ap_ohio}
        \end{subfigure}
        \vskip\baselineskip
        \begin{subfigure}[b]{0.45\textwidth}  
            \centering 
            \includegraphics{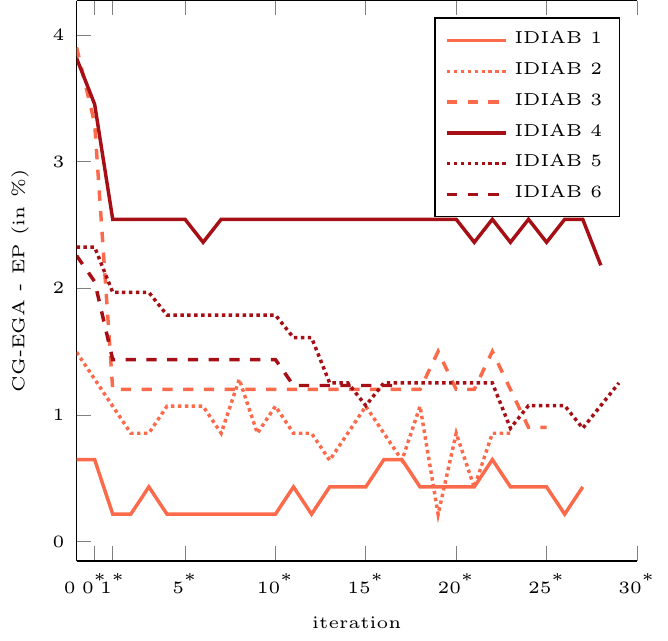}
            \caption[]%
            {{\small EP evolution of the IDIAB patients}}    
            \label{fig:cg_ega_algo_ep_idiab}
        \end{subfigure}
        \hfill
        \begin{subfigure}[b]{0.45\textwidth}  
            \centering 
            \includegraphics{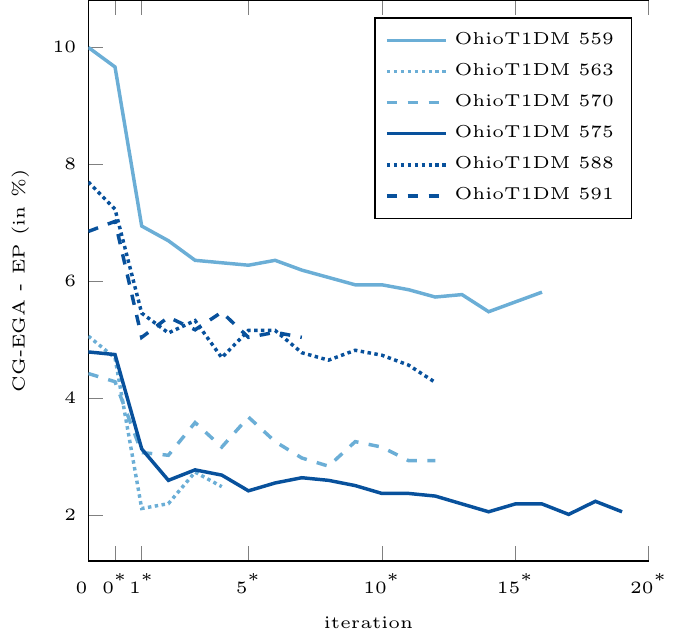}
            \caption[]%
            {{\small EP evolution of the OhioT1DM patients}}    
            \label{fig:cg_ega_algo_ep_ohio}
        \end{subfigure}
        \caption[ Evolution of the MASE and CG-EGA (AP and EP) metrics throughout the algorithm to progressively improve clinical acceptability for the IDIAB and OhioT1DM datasets. Iterations 0 and 0\textsuperscript{*} respectively represent the results of the model trained with the MSE cost function before and after smoothing the predictions.]
        {\small Evolution of the MASE and CG-EGA (AP and EP) metrics throughout the PICA algorithm for the IDIAB and OhioT1DM datasets. Iterations 0 and 0\textsuperscript{*} respectively represent the results of the model trained with the MSE cost function before and after smoothing the predictions.}
        \label{fig:metrics_iter}
\end{figure*}

Although we can conclude on the strength of using the gcMSE cost function in  the training of  deep models predicting future glucose levels in people with diabetes, the different results show us that there are many possible tradeoffs between accuracy and clinical acceptability. The PICA algorithm proposed in this study aims at selecting the best compromise between accuracy and clinical acceptability based on selection criteria. Figure \ref{fig:metrics_iter} gives a graphical representation of the changes in MASE, general AP rate and general EP rate of the models throughout the PICA algorithm for all the patients. As previously discussed, there is no clinical criterion for glucose predictive models yet, so the only criterion for stopping the algorithm here was the MASE exceeding 1. The figure first shows us that the number of iterations before stopping the algorithm is variable from one dataset to another, and also from one patient to another (25.0 $\pm$ 3.96 for the IDIAB dataset, and 11.66 $\pm$ 5.06 for the OhioT1DM dataset). This is explained first of all by the variable initial accuracy of the different patients, some patients being easier to predict than others (see iteration 0 on Figures \ref{fig:cg_ega_algo_mase_idiab} and \ref{fig:cg_ega_algo_mase_ohio}). As we have observed through the analysis of Table \ref{table:cgega_general_results}, the main improvements in clinical acceptability are made at the first iteration (iteration 1) of the algorithm when introducing the gcMSE cost function and exponential smoothing. Nevertheless, throughout the algorithm, the clinical acceptability gradually improves at the expense of the accuracy. We can see that the rate of deterioration and improvement is different from one patient to another, showing the very high inter-patient variability in the diabetic population.

Although there is currently no clinical criterion for glucose prediction models, we can analyze the use of two hypothetical criteria through Table \ref{table:cgega_thresholds}: a minimum AP rate, and a maximum EP rate. As expected, the harder the clinical criteria (higher threshold and/or combination of criteria), the lower the number of patients passing the clinical test. Only one patient in the IDIAB dataset managed to have simultaneously more than 97\% AP and less than 1\% EP. In addition, we can note a greater success of IDIAB patients on these clinical tests, compared to OhioT1DM patients. As previously mentioned, these differences in clinical performance are due to the difference in experimental systems. While the final evaluation of the OhioT1DM dataset is done every 5 minutes, it is done every 15 minutes for the IDIAB dataset. In addition, the glucose signal of IDIAB patients is overall less noisy, and therefore more stable and easier to predict. Thus, for a future practical use, the clinical criteria must be rigorously standardized.

Finally, we note that the MASE on the testing set (the one reported in Tables \ref{table:cgega_general_results} and \ref{table:cgega_cgega_results}) is slightly higher than 1 (1.03 and 1.01 for the IDIAB and OhioT1DM datasets). Using such a stopping criterion, we could have assumed that the final MASE on the testing set would be less than 1, as it is the case on the validation set. This happens because the test subset is not fully representative of the validation subset. This is due to the general small quantities of data in the datasets, negatively impacting the representativeness of these subsets. We also note that the standard deviation for the IDIAB dataset is higher, showing that the final value of the MASE is highly variable depending on the subject. Thus, the accuracy of the PICA algorithm would be improved by using more data (which would also improve the performance of the models in general).

% ResultsDatasetAllIter("pclstm","pclstm_iterative_gcmse_hypo_alliter",30,"ohio").number_of_patients_meeting_criteria([{"metric":"CG_EGA_AP","func": lambda x : np.greater_equal(x,0.95)},
%                                                                                                                  {"metric":"CG_EGA_EP","func": lambda x : np.less_equal(x,0.03)}])

\begin{table}[!ht]
\footnotesize
\centering
\caption{
Number of patients within a given dataset that can respect different clinical criteria (minimal AP rate or maximal EP rate) through the PICA algorithm.}
\label{table:cgega_thresholds}
\begin{center}

\begin{tabular}{c|c||c|c}
\toprule

\multicolumn{2}{c||}{\textbf{Clinical Criterion}} & \multicolumn{2}{c}{\textbf{Dataset}} \\
\midrule
AP ($\geq$) & EP ($\leq$) & IDIAB & Ohio \\
\midrule

80 & - & 6 & 6 \\
90 & - & 6 & 3 \\
95 & - & 4 & 0\\
97 & - & 3 & 0\\
\midrule
- & 7 & 6 & 6 \\
- & 5 & 6 & 4\\
- & 3 & 6 & 3 \\
- & 1 & 4 & 0\\
\midrule
80 & 7 & 6 & 6\\
90 & 5 & 6 & 3\\
95 & 3 & 4 & 0\\
97 & 1 & 2 & 0\\

\bottomrule
\end{tabular}
\end{center}
\end{table}{}
\section{Conclusion}

In this study, we proposed a framework for the integration of clinical criteria into the training of deep models. Clinical criteria are often different from standard statistical metrics used as loss functions. As a consequence the best model, given a loss function used during its training, is not necessarily the model with the best clinical acceptability. We address this issue from the perspective of the challenging task of predicting future glucose values of diabetic people.

In glucose prediction, the CG-EGA metric measures the clinical acceptability of the predictions. In particular, it assesses the safety of the predictions by looking at the prediction accuracy and the predicted rate of change accuracy. Moreover, the metric behaves differently for the different glycemic regions, some errors being more dangerous than others without being high amplitude errors. Starting from the cMSE loss function we proposed in an earlier work of ours \cite{de2019prediction} that penalizes the model during its training not only on prediction errors but also on predicted variation errors, we proposed to personalize the loss function to the glucose prediction task. Based on the CG-EGA, this personalization, called gcMSE, weights the errors differently depending on the scores obtained in the P-EGA and R-EGA grids. Finally, we proposed the PICA algorithm to obtain the solution that maximizes the accuracy of the model while at the same time respecting given clinical criteria.

We evaluate the different proposed loss functions and the PICA algorithm with two different diabetes datasets, the IDIAB and the OhioT1DM dataset. First, we showed that the cMSE loss function increase the coherence of successive predictions, improving the clinical acceptability of the models. However, this improvement comes at the cost of a decrease in the accuracy of the model. Then, we showed that the gcMSE further improves the clinical acceptability by reducing the rate of life-threatening errors. Finally, we demonstrate the usefulness of the PICA algorithm that help in choosing the desired tradeoff between general accuracy and clinical acceptability.

The analysis of different clinical criteria showed that not all the patients were able to meet them easily. It depends on the difficulty of the glucose prediction task of the patient, varying from patient to patient, but also on the nature of dataset, and in particular on the devices used for the data collection. These factors would need to be taken into account when creating future regulations for the use of such models by diabetic patients.

\section*{Acknowledgment}

This work is supported by the "IDI 2017" project funded by the IDEX Paris-Saclay, ANR-11-IDEX-0003-02. We would like to thank the French diabetes health network Revesdiab and Dr. Sylvie JOANNIDIS for their help in building the IDIAB dataset used in this study.

\bibliography{bibtex.bib}
\bibliographystyle{IEEEtran}

\ifCLASSOPTIONcaptionsoff
  \newpage
\fi

\end{document}